\documentclass{amsart}

\usepackage{epsfig}

\newtheorem{conjecture}{Conjecture}
\newtheorem{corollary}{Corollary}
\newtheorem{proposition}{Proposition}

\newcommand{\R}{\mathbb{R}}
\newcommand{\A}{{A}}
\newcommand{\E}[1]{\mathrm{e}^{\textstyle #1}}

\begin{document}

\title{On the Asymptotic Number of Plane Curves and Alternating Knots}

\author{Gilles Schaeffer}
\address{LIX--CNRS, Ecole Polytechnique, 91128 Palaiseau Cedex, France.}
\email{Gilles.Schaeffer@lix.polytechnique.fr\hfill\break\indent {\it Homepage:}
http://www.lix.polytechnique.fr/$\sim$schaeffe}
\author{Paul Zinn-Justin}
\address{LPTMS--CNRS, Universit\'e Paris-Sud, 91405 Orsay Cedex, France.}
\email{pzinn@lptms.u-psud.fr\hfill\break\indent {\it Homepage:}
http://ipnweb.in2p3.fr/lptms/membres/pzinn}

\date{04/03}

\begin{abstract}
  We present a conjecture for the power-law exponent in the asymptotic
  number of types of plane curves as the number of self-intersections
  goes to infinity.  In view of the description of prime alternating
  links as flype equivalence classes of plane curves, a similar
  conjecture is made for the asymptotic number of prime alternating
  knots.
  
  The rationale leading to these conjectures is given by quantum field
  theory. Plane curves are viewed as configurations of loops on a
  random planar lattices, that are in turn interpreted as a model of
  2d quantum gravity with matter. The identification of the
  universality class of this model yields the conjecture.
  
  Since approximate counting or sampling planar curves with more than
  a few dozens of intersections is an open problem, direct
  confrontation with numerical data yields no convincing indication on
  the correctness of our conjectures.  However, our physical approach
  yields a more general conjecture about connected systems of curves.
  We take advantage of this to design an original and feasible
  numerical test, based on recent perfect samplers for large planar
  maps.  The numerical data strongly support our identification with
  a conformal field theory recently described by Read and Saleur.  

\end{abstract}

\maketitle

\section{Introduction.}

\begin{figure}[t]
\begin{center}
\epsfig{file=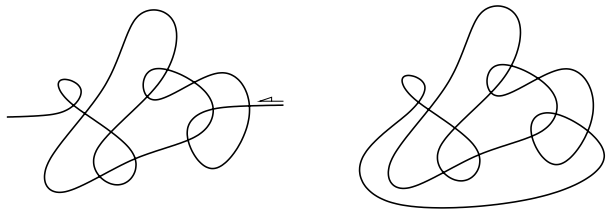, width=8cm}
\end{center}
\caption{An open plane curve and the associated closed  curve.
\label{open-close}}
\end{figure}

Our motivation for this work is the enumeration of topological
equivalence classes of smooth open and closed curves in the plane (see
Figure~\ref{open-close}; precise definitions are given in
Section~\ref{sec:doodles}).
The problem of characterizing closed curves was considered already by
Gauss and has generated many works since then: see \cite{rosenstiehl}
and references therein.  Our interest here is in the numbers $a_p$ and
$\alpha_p$ of such open and closed curves with $p$ self-intersections,
and more precisely we shall consider the asymptotic properties of
$a_p$ and $\alpha_p$ when $p$ goes to infinity. The numbers $a_p$ were
given up to $p=10$ in \cite{gz} and have been recently computed up
to $p=22$ by transfer matrix methods \cite{jzja}.  Asymptotically, as
$p$ goes to infinity, one expects the relation $a_p\sim4\,\alpha_p$ to
hold (see below), so that we concentrate on the numbers $a_p$.

In the present paper we propose a physical reinterpretation of the
numbers $a_p$ that leads to the following conjecture, and we present
numerical results supporting it.
\begin{conjecture}\label{main} 
There exist constants $\tau$ and $c$ such
that 
\[
\alpha_p\;\mathop{\sim}_{p\to\infty}\;\frac14\,a_p
\;\mathop{\sim}_{p\to\infty}\; c\,\tau^p\cdot p^{\gamma-2},%\cdot\log^\nu p
\]
where 
\begin{equation}\label{critical-exp}
\gamma=-\frac{1+\sqrt{13}}6\;\doteq\;-0.76759...
\end{equation}
\end{conjecture}
{}From the data of \cite{jzja} one has the numerical estimate:
$\tau\doteq 11.4$.  But the main point in Conjecture~\ref{main} lies
not so much in the existence of $\tau$ 
% GS: je pense qu'on pourrait ajouter ici:
% (that follows from subadditivity arguments)
% mais il faudrait verifier. 
as in the explicit value of $\gamma$.  It should indeed be observed
that $\gamma$, rather than $\tau$, is the interesting information in
the asymptotic of $a_p$. Observe for instance that the value of
$\gamma$ is left unchanged if one redefines the size of a closed curve
as the number $p'=2p$ of arcs between crossings. More generally, as
discussed in Section~\ref{sec:conclusion}, the exponent
$\gamma$ determines the branching behavior of generic large curves
under the uniform distribution, and is \emph{universal} in the sense
that the same value is expected in the asymptotic of related families
of objects like prime self-intersecting curves or alternating knots.

Conjecture~\ref{main} is similar in nature to the conjecture of Di
Francesco, Golinelli and Guitter on the asymptotic behavior of the
number of plane meanders \cite{dFGG}.  The two problems do not fall
into the same universality class (in particular the predictions for
the exponent $\gamma$ are different in the two problems). However our
approach to design a numerical test is applicable also to the meander
problem.

The rest of the paper is organized as follows. Precise definitions are
given, and a more general family of drawings is introduced that play
an important r\^ole in the identification of the associated physical
model (Section~\ref{sec:maps}).  The physical background leading to
the conjecture is then discussed (Section~\ref{sec:gravity}) and a
numerically testable quantity is proposed
(Section~\ref{sec:parameter}).  The sampling method is briefly
presented (Section~\ref{sec:sampling}) before the analysis of
numerical data (Section~\ref{sec:simulations}).  We conclude with
some variants and corollaries of the conjecture
(Section~\ref{sec:conclusion}).

\section{Plane curves and colored planar maps}\label{sec:maps}
\subsection{Plane curves and doodles} \label{sec:doodles}

For $p$ a positive integer, let $\mathcal{A}_p$ be the set of
equivalence classes of self-intersecting loops $\gamma$ in the plane,
that is: (i) $\gamma$ is a smooth mapping $S^1\to\R^2$; (ii) there are
$p$ points of self-intersection, all of which are regular crossings;
(iii) two loops $\gamma$ and $\gamma'$ are equivalent if there exists
an orientation preserving homeomorphism $h$ of the plane
% une variante possible:
%and an homeomorphism $\varepsilon$ of $S^1$ such that
%$\gamma'=h\circ\gamma\circ\varepsilon$.
such that $\gamma'(S^1)=h(\gamma(S^1))$.

Similarly let $A_p$ be the set of equivalence classes of
self-intersecting open curves $\gamma$ in the plane: (i) $\gamma$ is a
smooth mapping $[0,1]\to\mathbb{R}^2$ and $\gamma(0)$ and $\gamma(1)$
belong to the infinite component of
$\mathbb{R}^2\setminus\gamma((0,1))$; (ii) there are $p$ points of
self-intersection, all of which are regular crossings; (iii) two open
curves are equivalent if there exists an orientation preserving
homeomorphism $h$ of the plane 
% la meme variante pour les courbes ouvertes:
%and an homeomorphism $\varepsilon$ of $[0,1]$ fixing $0$ and $1$ 
%such that $\gamma'=h\circ\gamma\circ\varepsilon$.
such that $\gamma'([0,1])=h(\gamma([0,1]))$ and $\gamma'(i)=h(\gamma(i))$ for $i=0,1$.

Observe that, unlike closed curves, open curves are oriented from the
initial point $\gamma(0)$ to the final point $\gamma(1)$.  Moreover a
unique closed curve is obtained from an open curve by connecting the
final point to the initial one in counterclockwise direction around
the curve. These definitions are illustrated by
Figure~\ref{open-close}.

\medskip

\begin{figure}[t]
\begin{center}
\epsfig{file=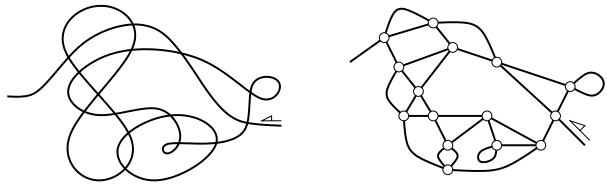, width=8cm}
\end{center}
\caption{A doodle and the associated rooted planar map.
\label{doodle}}
\end{figure}

In order to study the families $\mathcal{A}_p$ and $A_p$ and to obtain
Conjecture~\ref{main} we introduce a more general class of drawings,
that we call \emph{doodles}. For given positive integers $p$ and $k$,
let ${A}_{k,p}$ be the set of equivalence classes of $(k+1)$-uples
$\Gamma=(\gamma_0,\gamma_1,\ldots,\gamma_k)$ of curves drawn on the
plane such that: (i) the curve $\gamma_0$ is an open curve of the
plane: $\gamma_0$ is a smooth mapping $[0,1]\to\R^2$, and
$\gamma_0(0)$ and $\gamma_0(1)$ belong to the infinite component of
$\R^2\setminus(\gamma_0((0,1))\cup\bigcup_i\gamma_i(S^1))$; (ii) for
$i\geq1$, each $\gamma_i$ is a loop, that is a smooth map
$S^1\to\R^2$; (iii) there are $p$ points of intersection (including
possibly self-intersections) of these curves, all of which are regular
crossings; (iv) the union of the curves is connected, (v) two doodles
$\Gamma=(\gamma_0,\ldots,\gamma_k)$ and
$\Gamma'=(\gamma_0',\ldots,\gamma'_k)$ are equivalent if there exists
an orientation preserving homeomorphism $h$ of the plane such that
$\gamma'_0([0,1])\cup\bigcup_i\gamma'_i(S^1)=
h(\gamma_0([0,1])\cup\bigcup_i\gamma_i(S^1))$ and
$\gamma'_0(x)=h(\gamma_0(x))$, for $x=0,1$.
%, a permutation $\sigma$ of $\{1,\ldots,k\}$ and $k$ homeomorphisms of
%${S}^1$ such that
%$(\gamma_0',\gamma_1',\ldots,\gamma_k')=(h\circ\gamma_{0},
%h\circ\gamma_{\sigma(1)}\circ\varepsilon_1,\ldots,
%h\circ\gamma_{\sigma(k)}\circ\varepsilon_k)$. 
In other terms, a doodle is made of an open curve intersecting a set
of loops, that are considered up to continuous deformations of the
plane. An example of doodle is given in Figure~\ref{doodle} (left-hand side).

\subsection{Colored planar maps}
An equivalent presentation of doodles is in terms of rooted {planar
  maps} \cite{tutte,tutte2}. A \emph{planar map} is a proper embedding
of a connected graph in the plane considered up to homeomorphisms of
the plane.  It is \emph{4-regular} if all vertices have degree four.
It is \emph{rooted} if one {root edge} is marked on the infinite face
and oriented in counterclockwise direction. Equivalently the root edge
can be cut into an in- and an out-going half-edge (also called
\emph{legs}) in the infinite face.  There is an immediate one-to-one
correspondence between doodles with $p$ crossings, and 4-regular
planar maps with $p$ vertices and two legs.  This correspondence is
illustrated on Figure~\ref{doodle}. 

We shall consider the number $a_{k,p}=\mathrm{card}\,{A}_{k,p}$ of
doodles with $p$ crossings and $k$ loops and more specifically we
shall consider the asymptotic properties of $a_{k,p}$ as $p$ (and
possibly $k$) goes to infinity.   It turns out to be convenient to
introduce the generating function $a_p(n)$ as $k$ varies:
\begin{equation}
a_p(n)=\sum_{k=0}^\infty a_{k,p}\, n^k
\end{equation}
The requirement that a doodle is connected implies that it cannot
contain more loops than crossings so that $a_p(n)$ is a polynomial in
the (formal) variable\footnote{The subsequent interpretation in terms
  of colored doodles and the strong tradition in the physics litterature are
  our (admittingly poor) excuses for the use of $n$ to denote a formal
  variable.}  $n$.  For real valued $n$, $a_p(n)$ can be understood as
a weighted summation over all doodles with $p$ crossings, and, more
specifically for $n$ a positive integer, $a_p(n)$ can be interpreted
as the number of \emph{colored} doodles in which each loop has been
drawn using a color taken from a set of $n$ distinct colors.

On the one hand, in the special case $k=0$, $a_{0,p}=a_p$ gives by
definition the number of open self-intersecting plane curves. Observe
that $a_p$ is also given by $n=0$ since $a_p(0)=a_{0,p}$. On the other
hand, the generating functions of the $a_p(n)$ for other values of
$n$, namely $n=1,2,-2$, have been computed exactly (see respectively
\cite{tutte,zjzb,pzjb}).  We elaborate now on the case $n=1$ since it
will play a crucial role in what follows.

The number $a_p(1)$ counts the number of doodles with $p$ crossings
irrespective of the number of loops $k$. In terms of maps, $a_p(1)$ is
the number of rooted 4-regular planar maps with $p$ vertices.  The
number of such planar maps is known \cite{bipz,tutte}, from which one
can compute the asymptotics:
\begin{equation}\label{one}
a_p(1)=2{3^p\,(2p)!\over p!(p+2)!}
\;\mathop{\sim}_{p\to\infty}\;
%\buildrel p\to \infty \over\sim 
\frac{2}{\sqrt{\pi}} \;12^p p^{-5/2}.
\end{equation}
Observe in this case the power-law exponent $-5/2$, which is
\emph{universal} for rooted planar maps in the sense that it is
observed for a variety of families of rooted planar maps (see
\cite{gao}). As opposed to this, the exponential growth factor $12$ is
specific to the family of rooted 4-regular planar maps.

There is a physical interpretation of the power-law behavior
$p^{-5/2}$: it is given by two-dimensional gravity. This explanation
begs to be generalized to any $n$, and we shall explore such a
possibility now.

\section{Two-dimensional quantum gravity and 
asymptotic combinatorics}\label{sec:gravity}
% 2D quantum gravity and asy combi
The purpose of this section is to give the rationale behind our
conjectures.  We place the discussion at a rather informal level that
we hope achieves a double purpose: on the one hand it should give an
intuition of the path leading to our conjectures to the reader with
zero-knowledge in quantum field theory (QFT), and on the other hand it
should convince the expert of this (quantum) field.  For our defense,
let us observe that filling in more details would require a complete
course on QFT, with the result of not getting much closer to a
mathematical proof.

\subsection{From planar maps to two-dimensional quantum gravity}
The main idea of the physical interpretation of the numbers $a_p(1)$
is to consider planar maps as discretized random surfaces (with the
topology of the sphere%; see Figure~\ref{fig:sphere}
).  As the number
of vertices of the map grows large, the details of the discretization
can be assimilated to the fluctuations of the {\it metric}\/ on the
sphere. To make this idea more precise, let us describe a way to
associate a metric on the sphere to a given 4-regular map $m$: to each
vertex of $m$ associate a unit square and identify the sides of these
squares according to the edges of $m$ (arbitrary number of corners of
squares get identified); the result is by construction a metric space
with the topology of a sphere. Upon taking a 4-regular map uniformly
at random in the set of map with $p$ edges, a random metric sphere
with area $p$ is obtained.

%\begin{figure}
%\caption{A figure of a map on the sphere, to be made} 
%\label{fig:sphere}
%\end{figure}

Now, physics tells us that the metric is the dynamical field of
general relativity \emph{i.e.}\ gravity, and that this type of
fluctuations in the metric are characteristic of a quantum theory.  In
our case it means that, as $p$ becomes large, the discrete nature of
the maps can be ignored and there exists a scaling limit, the
properties of which are described by {two-dimensional euclidian
  quantum gravity}. In particular any parameter of random planar maps that
makes sense in the scaling should converge to its continuum analog.
A fundamental parameter of this kind turns out to be precisely the
number of (unrooted) planar maps: it is expected to scale to the
partition function $Z_{\rm g}(A)$ of two-dimensional quantum gravity with
spherical topology at fixed area $A$, through a relation of the form
\begin{equation}\label{puregravity}
\frac1pa_p(1)\;\mathop{\sim}_{p\to\infty}\; Z_{\rm  g}(\A), \textrm{ with }
\A \textrm{ proportional to }p.
\end{equation} 
(Here the factor $1/p$ is due to the fact that the partition function
does not takes the rooting into account.)  The only thing we want to
retain from $Z_{\rm g}(A)$ is that the power law dependance of its large
area asymptotic takes the form $A^{-7/2}$, in accordance with
Formula~(\ref{one}).  (Trying to give here a precise description of
the partition function $Z_{\rm g}$ would carry us to far away, and anyway
the arguments in this section are non rigourous.)
\medskip

In the case $n=1$, this is the whole physical picture: a fluctuating
but empty two-dimensional spacetime -- there is no matter in it. What
happens when $n\ne 1$? As already discussed, an appealing image is to
consider that one must ``decorate'' the planar map by coloring each
curve $\gamma_i$ with $n$ colors.  Alternatively, the physicist's view
would be to consider that we have put a statistical lattice model (of
crossing loops) on a random lattice (the planar map).  This perfectly
fits with the previous interpretation of the planar map as a
fluctuating two-dimensional spacetime: as we learn from physics, in
the limit of large size, adding a statistical lattice model amounts to
{coupling matter to quantum gravity}. Matter is described by a quantum
field theory (QFT) living on the two-dimensional spacetime.  The
parameters of the lattice model that survive in the scaling limit are
recovered in the critical (long distance) behavior of this QFT, which
in turn is described by a conformal field theory (CFT).  Then,
provided we can find a CFT describing the lattice model corresponding
to a given $n\neq1$, the analog of Relation~(\ref{puregravity}) holds
with the partition function $Z_{{\rm g+CFT}(n)}(\A)$ of this CFT coupled to
gravity: in the large size limit,
\begin{equation}\label{mattergravity}
\frac1pa_p(n)\;\mathop{\sim}_{p\to\infty}\; Z_{{\rm  g+CFT}(n)}(\A).
\end{equation} 

In this picture, the only thing we need to know about the CFT that
describes the scaling limit of our model is its \emph{central charge}
$c$, which roughly counts its number of degrees of freedom. Indeed,
the study of CFT coupled to gravity was performed in
\cite{kpz2,kpz3,kpz1}, resulting in the following fundamental
prediction: the partition function $Z_{\rm g+CFT}(\A)$ of gravity
dressed with matter has a power-law dependence on the area of the form
$\A^{\gamma-3}$ where the critical exponent $\gamma$ depends only on
the central charge of the underlying CFT via (for $c<1$)
\begin{equation}\label{gam}
\gamma={c-1-\sqrt{(1-c)(25-c)}\over 12}.
\end{equation}

Returning to our asymptotic enumeration problem (not forgetting the
extra factor $p$ which comes from the marked edge), we find:
\begin{equation}\label{grav}
a_p(n)\; \mathop{\sim}_{p\to \infty}\;
\E{\sigma\,p+(\gamma-2)\, \log p+\kappa}
\end{equation}
where $\sigma$, $\kappa$ are unspecified ``non-universal'' parameters,
whereas the ``universal'' exponent $\gamma$ is given by
Eq.~(\ref{gam}) with the central charge $c$ of the {\it a priori}\/
unknown underlying CFT$(n)$.  The absence of matter,
that is the case $n=1$, corresponds to a CFT with central charge $c=0$: one
recovers $\gamma-2=-5/2$ as expected from Eq.~(\ref{one}).
In general, all parameters in Eq.~(\ref{grav}) are functions of $n$;
assuming furthermore that their dependence on $n$ is smooth in a
neighborhood of $n=1$, one can recover by Legendre transform of $\sigma(n)$
the asymptotics of $a_{k,p}$ as $k$ and $p$ tend to infinity with the
ratio $k/p$ fixed.
Observe finally that the knowledge of the CFT could give more
informations. For instance, the irrelevant operators of the CFT control
subleading corrections to Eq.~(\ref{grav}).

\subsection{The identification of two candidate models}\label{sec:candidates}
We now come to the issue of the determination of the CFT for a
arbitrary $n$.  An observation made in \cite{pzj}, based on a matrix
integral formulation, is that this CFT must have an $O(n)$ symmetry
(for $n$ positive integer -- for generic $n$ this symmetry becomes
rather formal and cannot be realized as a unitary action of a compact
group on the Hilbert space).  There exists a well-known statistical
model with $O(n)$ symmetry, a model of (dense) {\it non-crossing}\/
loops \cite{ni}, whose continuum limit for $|n|<2$ is described by a
CFT with central charge
\begin{equation}\label{cfirst}
\hskip2cm 
c_{\rm I}=1-6(\sqrt{g}-1/\sqrt{g})^2
\hskip1cm
n=-2\cos(\pi g),\quad 0<g<1
\end{equation}
In \cite{pzj} it was speculated that there is no phase transition
between the model of non-crossing loops, which we call model I, and
our model of crossing loops, and therefore the central charge is the
same and given by Eq.~(\ref{cfirst}). If this were the case, the study
of irrelevant operators of this CFT would allow moreover to predict
that subleading corrections to Eq.~(\ref{grav}) have power-law
behavior for all $|n|<2$ with exponents depending continuously on $n$.

However, another scenario is possible. In \cite{rs}, it was suggested
that $O(n)$ models, for $n<2$, possess in general a low temperature
phase with {\it spontaneous symmetry breaking} of the $O(n)$ symmetry
into a subgroup $O(n-1)$. This is a well-known mechanism\footnote{The
  Mermin--Wagner theorem, which forbids spontaneous symmetry breaking
  of a continuous symmetry in two dimensions, only applies to $n$
  integer greater or equal to $2$.} in QFT (see e.g.\ \cite{zj}
chapters 14, 30), which produces Goldstone bosons living on the
homogeneous space $O(n)/O(n-1)=S^{n-1}$.  In the low energy limit the
bosons become free and the central charge is simply the dimension of
the target space $S^{n-1}$:
\begin{equation}\label{csec}
c_{\rm II}=n-1\hskip 1cm n<2
\end{equation}
For generic real $n$ ($n<2$) this is only meaningful in the sense of
analytic continuation, but we assume it can be done and call it model
II. This CFT possess a marginally irrelevant operator, leading to main
corrections to leading behavior (\ref{grav}) of {\it logarithmic} type
\emph{i.e.}\ in $\frac{1}{\log p}$, $\frac{\log\log p}{(\log p)^2}$
etc.

It was furthermore argued in \cite{rs} that the critical phase of
model II is generic in the sense that the low-energy CFT is not
destroyed by small perturbations -- the most relevant $O(n)$-invariant
perturbation is the action itself, which corresponds to a marginally
irrelevant operator for $n<2$.  On the contrary, the model I of
non-crossing loops is unstable to perturbation by crossings; some
numerical work on regular lattices (at $n=0$) \cite{jrs} tends to
suggest that it flows towards model II.

Note that both Conjectures (\ref{cfirst}) and (\ref{csec})
supply the correct value $c=0$ for $n=1$ and
$c=1$ for the limiting case $n=2$.\footnote{Actually,
the two resulting $c=1$ theories are not identical:
the one from model I seems to be the wrong one,
although this is a subtle point on which we do not dwell here.}
Of course, in no way do we claim
that these are the {\it only}\/ possible scenarios which fit with
known results -- one might have a plateau
of non-critical behavior ($c=0$) around $n=1$, for instance;
or two-dimensional quantum gravity universality arguments
might not apply at all in some regions of $n$ -- but
they seem the most likely candidates and therefore it is important
to find a numerically accessible quantity which at least
discriminates between the two conjectures.

\section{The general conjectures and a testable parameter}\label{sec:parameter}
The physical reinterpretation of doodles as a model on random planar
lattices has led us to postulate that the weighted summation over
doodles satisfies
\[
a_p(n)\sim c_0(n)\,\tau(n)^p\cdot p^{\gamma(n)-2},
\]
with the critical exponent $\gamma(n)$ given in terms of the central charge $c(n)$ by
\[
\gamma(n)=\frac{c(n)-1-\sqrt{(1-c(n))(25-c(n))}}{12}.
\]
Moreover we have presented two concurrent models which fix the value of
$c(n)$. Since negative values of $n$ create additional technical difficulties
(appearance of complex singularities in the generating function, cf \cite{pzjb}),
we formulate the conjectures in the restricted range $0\le n<2$:

\begin{conjecture}[Model I]\label{model1}
Colored doodles are in the universality class of dense non-crossing loops,
so that for $0\le n<2$, $n=-2\cos(\pi g)$, $1/2\le g<1$,
\[ \label{conj1}
c(n)=1-6(\sqrt g-1/\sqrt g)^2.
\]
\end{conjecture}

\begin{conjecture}[Model II]\label{model2}
Colored doodles are in the universality class of models with spontaneously
broken $O(n)$ symmetry, so that for $0\le n<2$, 
\[ \label{conj2}
c(n)=n-1.
\]
\end{conjecture}
Observe that Conjecture~\ref{model2} implies Conjecture~\ref{main} for
$n=0$, while Conjecture~\ref{model1} would give
$c(0)=1-6(\sqrt2-1/\sqrt2)^2=-2$ and $\gamma(0)=-1$.
According to the discussion of the previous section,
Conjecture~\ref{model2} appears more convincing. In order to get a
numerical confirmation, we look for a way to discriminate between
the two.

Since the model at $n=1$ is much easier to manipulate, we look
for such a quantity at $n=1$. Of course the known value of the
exponent $\gamma(1)$ is a natural candidate but as already mentioned
both conjectures agree on this: we propose instead the derivative of
the exponent at $n=1$,
\begin{equation}
\gamma'\equiv {d\over d n}_{|n=1} \gamma(n).
\end{equation}
The reason that it can easily be computed numerically is that it
appears in the expansion of the average number of loops $\left< k
\right>_p$ for a uniformly distributed random planar map with $p$
vertices. Indeed one easily finds
\begin{equation}\label{defk}
\left<k\right>_p={d\over dn}_{|n=1} \log a_p(n)
\mathop{=}_{p\to \infty} \sigma' p + \gamma' \log p + \kappa'
+o(1)
\end{equation}
Here we have assumed expansion (\ref{grav}) to be uniform with
smoothly varying constants $\sigma(n)$, $\gamma(n)$, $\kappa(n)$ in
some neighborhood of $n=1$, and written 
$\sigma'\equiv {d\over d n}_{|n=1} \sigma(n)$,
$\kappa'\equiv {d\over d n}_{|n=1} \kappa(n)$. 

The conjectures \ref{conj1} and \ref{conj2} provide
the following predictions for $\gamma'$:
\begin{equation}\label{thetest}
\gamma'=\begin{cases}
\frac{3\sqrt{3}}{4\pi}= 0.413\ldots& \hbox{in CFT I}\cr
\frac{3}{10}=0.3&\hbox{in CFT II}\cr\end{cases}
\end{equation}

The quantity $\left<k\right>_p$ is not known theoretically, so that we
cannot immediately conclude in either direction. However it is
possible to estimate it numerically using random sampling. 

\section{Sampling random planar maps}\label{sec:sampling}
In this section we present the algorithm we use to sample a random map
from the uniform distribution on rooted 4-regular planar maps with $p$
vertices.
The problem of sampling random planar maps with various constraints
under the uniform distribution was first approached in mathematical
physics using markov chain methods \cite{kazakov,
ambjornetal}. However these methods require a large and unknown number
of iterations, and only approximate the uniform distribution.  Another
approach was proposed based on the original recursive decompositions
of Tutte \cite{tutte} but has quadratic complexity \cite{migdal}, and
is limited as well to $p$ of order a few thousands.

We use here a more efficient method that was proposed in
\cite{S97,S99} along with a new derivation of Tutte's formulas.  The
algorithm, which we outline here in the case of 4-regular maps,
requires only $O(p)$ operations to generate a map with $p$ vertices
and manipulates only integers bounded by $O(p)$. Moreover maps are
sampled exactly from the uniform distribution. The only limitation
thus lies in the space occupied by the generated map. In practice we
were able to generate maps with up to 100 million vertices, with a
generation speed of a million vertices per second.

\begin{figure}
\begin{center}
\epsfig{file=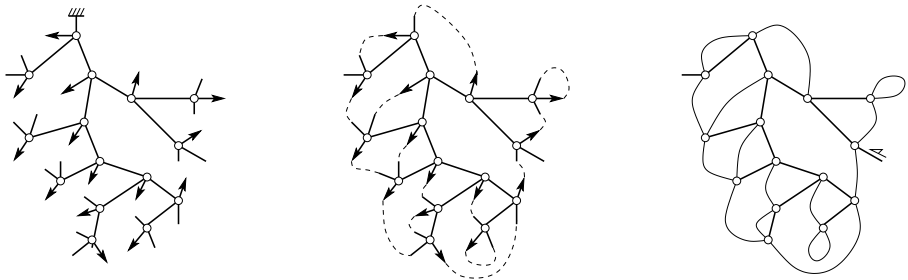, width=12cm}
\end{center}
\caption{A blossom tree and its closure. Buds are represented by
arrows. Dashed edges connect pairs of matched buds and leaves.
\label{tree}}
\end{figure}

The algorithm relies on a correspondence between rooted 4-regular
planar maps and a family of trees that we now define.  A \emph{blossom
  tree} is a planted plane tree such that
\begin{itemize}
\item vertices of degree one are of two types: buds and leaves;
\item each inner vertex has degree four and is incident to exactly one
  bud;
\item the root is a leaf.
\end{itemize}
An example of blossom tree is shown on Figure~\ref{tree}.
By definition a blossom tree with $p$ inner vertices has $p+2$ leaves
(including the root) and $p$ buds. Observe that removing the buds of a
blossom tree gives a planted complete binary tree with $p$ inner
vertices, and that conversely $3^p$ blossom trees can be constructed
out of given binary tree with $p$ inner vertices. Since the number of
binary trees with $p$ inner vertices is well known to be the Catalan number 
$\frac{1}{p+1}{2p\choose p}$, 
the number of blossom tree is seen to be
\[
3^p\cdot\frac1{p+1}{2p\choose p}.
\]

Let us define the \emph{closure} of a blossom tree. An example is
shown on Figure~\ref{tree}. Buds and leaves of a blossom tree with $p$
inner vertices form in the infinite face a cyclic sequence with $p$
buds and $p+2$ leaves. In this sequence each pair of consecutive bud
and leaf (in counterclockwise order around the infinite face) are
merged to form an edge enclosing a finite face containing no unmatched
bud or leaf. Matched buds and leaves are eliminated from the sequence
of buds and leaves in the infinite face and the matching process can
be repeated until there is no more buds available. Two leaves then
remain in the infinite face.
 
\begin{proposition}[\cite{S97}]
  Closure defines a $(p+2)$-to-$2$ correspondence between blossom trees
  and rooted four-regular planar maps. In particular the number of 
  rooted four-regular planar maps is 
  \[
  \frac{2}{p+2}\cdot\frac{3^p}{p+1}{2p\choose p}.
  \]
\end{proposition}

This proposition implies that to generate a random map according to
the uniform distribution on rooted 4-regular planar maps with $p$
vertices one can generate a blossom tree according to the uniform
distribution on blossom tree and apply closure.
A synopsis of the sampling algorithms is given below. An implementation is
available on the web page of G.S.

\medskip
\vbox{
\hrule
\medskip

\noindent\textsf{Random sampling of a rooted 4-regular maps with $p$ vertices.} 

\emph{Step 1.} Generate a random complete binary tree $T_1$ according
to the uniform distribution on complete binary trees with $p$ inner
vertices. (This is done in linear time using \emph{e.g.}\ prefix
codes \cite{knuth}.)
  
\emph{Step 2.} Convert $T_1$ into a random blossom tree $T_2$ from the
uniform distribution on blossom trees with $p$ inner vertices:
independantly add a bud on each vertex in a uniformly chosen position
among the three possibilities.

\emph{Step 3.} Use a stack (a.k.a.\ a last-in-first-out waiting line) to
realise the closure of $T_2$ in linear time: Perform a
counterclockwise traversal of the infinite face until $p$ buds and
leaves have been matched; when a bud is met, put $b$ into the
stack; when a leaf $\ell$ is met and the stack is non empty,
remove the last bud entered in the stack and match it with $\ell$.
  
\emph{Step 4.} Choose uniformly the root between the two remaining
leaves.

\medskip\hrule\medskip
}

\section{Simulation results}\label{sec:simulations}
% simulation results
%
The algorithm described in the previous section allows to generate
random rooted 4-regular planar maps with $p$ vertices and two legs, with uniform
probability. One can compute various quantities related to the map thus
generated and then average over a sample of maps, as always in
Montecarlo simulations. Here the main quantity of interest is
the number of loops of the map. If we generate $N$ maps of size $p$
so that the $i$th %^{\rm th}$ 
map has $k_{p,i}$ loops, $1\le i\le N$, 
then ${1\over N}\sum_{i=1}^N k_{p,i}$ 
has an expectation value of $\langle k \rangle_p$ 
and a variance of ${1\over N} \langle\langle k^2\rangle\rangle_p$,
where $\langle k \rangle_p = {d\over d n}_{|n=1} \log a_p(n)$
and $\langle\langle k^2 \rangle \rangle_p = {d^2\over d n^2}_{|n=1} \log a_p(n)$
(the latter can of course itself be estimated as the expectation value
of ${1\over N-1} \sum_{i=1}^N k_{p,i}^2-{1\over N(N-1)} (\sum_{i=1}^n k_{p,i})^2$).
According to Eq.~(\ref{grav}), both $\langle k \rangle_p$ and $\langle\langle
 k^2 \rangle\rangle_p$ are of order $p$ for $p$ large. However, we are interested
in corrections to the leading behavior
of $\langle k \rangle_p$ which are %essentially of order $1$, 
of order $\log p$,
cf Eq.~(\ref{defk}), so that we need to keep the {\it absolute}\/ 
error small.
This implies that the size of the sample $N$ should scale like $p$, or that the
computation time grows quadratically as a function of $p$.

In practice we have produced data for $p=2^\ell$ with $\ell\le 24$, 
the sample size being
of the order of up to $10^7$. To ensure a good sampling we used the ``Mersenne
twister'' pseudo-random generator \cite{mersenne}, 
which is both fast and unbiased. The last
few values of $\ell$ are only given to show where the statistical error begins
to grow large due to limited memory and computation time. 
Let us call $k_\ell$ the numerical value found for $\langle k\rangle_{p=2^\ell}$.
The results obtained are shown on table \ref{tab:data}.

% careful! open curve not counted -> shift of 1
%
\begin{table}
\caption{Numerical values $k_\ell$ of the average number of loops
of maps with $p=2^\ell$ vertices. The error (standard deviation) on the last digit is given in parentheses.}
\begin{tabular}{ccccccccc}
\hline
$\ell$&1&2&3&4&5&6\\
$k_\ell$&0.1111(0)&0.3228(0)&0.6605(0)&1.2120(0)& 2.1640(1)&3.8970(1)\\
\hline
$\ell$&7&8&9&10&11&12\\
$k_\ell$&7.1764(1)&13.5372(1)&26.0524(2)&50.8704(2)&100.2890(3)&198.9060(6)\\
\hline
$\ell$&13&14&15&16&17&18\\
$k_\ell$&395.916(1)&789.716(2)&1577.089(4)&3151.607(7)&6300.44(1)&12597.83(2)\\
\hline
$\ell$&19&20&21&22&23&24\\
$k_\ell$&25192.45(3)&50381.35(5)&100759.0(1)&201514.3(2)&403023.8(4)&806043.2(7)\\
\hline
\end{tabular}
\label{tab:data}
\end{table}

%Due to the excellent accuracy of the results, various methods can be used
%to analyze these data. For example,
%one can extract the coefficients of the asymptotic expansion (\ref{defk}) by 
%performing appropriate combinations of the $k_\ell$
%to cancel all terms of the expansion but one and extrapolating the result.
%Our main interest lies in the value of $\gamma'$.
%Defining $u_\ell=2k_\ell-k_{\ell+1}$, one then compares $u_\ell$ with the expected
%asymptotic behavior
%$u_\ell=(\ell-1)\gamma' \log 2 + \kappa'+O(1/\ell)$. 
%The result is shown on Fig.~\ref{fit}.

\begin{figure}[ht]
\begin{center}
\epsfig{file=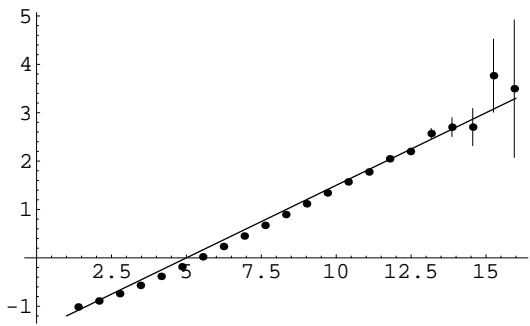, width=7cm}
\end{center}
\caption{The set of points $u_\ell=2k_\ell-k_{\ell+1}$ as a function of
$\log p = \log 2 \cdot \ell$ with their error bars, as well as a
proposed asymptote of slope $0.3$.
\label{fit}}
\end{figure}

First, as a rough check of the asymptotic behavior, let us define
$u_\ell=2k_\ell-k_{\ell+1}$. If expansion (\ref{defk}) is correct, then
$u_\ell$ must display an affine behavior as a function of $\ell$:
$u_\ell=(\ell-1)\gamma' \log 2 + \kappa'+O(1/\ell)$. Indeed, as one can
see on Fig.~\ref{fit}, this is the case. 

By comparison with the proposed asymptote 
it seems clear that $\gamma'$ is close to $0.3$. To make this statement
more precise, one can try to fit the set of the $k_\ell$ to
$\sigma' p+\gamma' \log p+\kappa'$, where $\ell$ ranges
from $\ell=\ell_{min}$
to $\ell=\ell_{max}=24$ and $\ell_{min}$ is varied.
The results are reported on Tab.~\ref{morefits}. 
Unfortunately the confidence
level remains fairly low until $\ell_{min}$ becomes so high that statistical error
is huge, which tends to indicate strong corrections to the proposed fit.

\begin{table}
\caption{Fits for the $k_\ell$. $\chi^2$ is the minimized weighted 
sum of squared errors.}
\begin{tabular}{cccccccc}
\hline
$\ell_{min}$ & 2 & 3 & 4 & 5 & 6 & 7\\
$\sigma'$ & 0.04804410 & 0.04804398 & 0.04804388 & 0.04804382 & 0.04804377 & 0.04804374\\
$\gamma'$ &0.2952 &0.3018 &0.3071 &0.3113 &0.3148 &0.3175\\
$\kappa'$ & -0.364 & -0.408 & -0.445 & -0.475 & -0.501 & -0.522\\
$\chi^2$ & 18273 & 8067.63 & 3414.53 & 1384.07 & 522.297 & 187.471\\
\hline
$\ell_{min}$ & 8 & 9 & 10 & 11 & 12 & 13\\
$\sigma'$ & 0.04804371 & 0.04804370 & 0.04804369 & 0.04804368 & 0.04804368 & 0.04804367\\
$\gamma'$ &0.3196 &0.3213 &0.3226 &0.3236 &0.3246 &0.3266\\
$\kappa'$ & -0.539 & -0.553 & -0.563 & -0.572 & -0.582 & -0.600\\
$\chi^2$ & 64.4297 & 24.3678 & 12.7841 & 9.30634 & 8.00342 & 6.30457\\
\hline
$\ell_{min}$ & 14 & 15 & 16 & 17 & 18 & 19\\
$\sigma'$ & 0.04804366 & 0.04804365 & 0.04804364 & 0.04804364 & 0.04804363 & 0.04804365\\
$\gamma'$ &0.3289 &0.3340 &0.3440 &0.3392 &0.3700 &0.3129\\
$\kappa'$ & -0.624 & -0.680 & -0.795 & -0.737 & -1.13 & -0.373\\
$\chi^2$ & 5.69736 & 4.72577 & 3.66152 & 3.58534 & 2.8422 & 2.24532\\
\hline
\end{tabular}
\label{morefits}
\end{table}

It is important to understand that if Conjecture 2 were true, then the
corrections to asymptotic behavior would be power-law -- starting with $p^{-1/2}$.
This means that the procedure used 
in Tab.~\ref{morefits} should converge quickly to the correct
values of $\sigma'$, $\gamma'$, $\kappa'$ (to check this we have performed
a similar analysis with a model of non-crossing loops on random planar maps
and obtained fast convergence with high accuracy -- 2 digits on $\gamma'$).
Here %the convergence is slow, and 
the range of values 
of $\gamma'$ seems to be $0.29$--$0.34$, far from the value predicted
by Conjecture 2. It is therefore our view that
the numerical data render Conjecture 2 extremely unlikely.

On the other hand, the value $0.3$ predicted by Conjecture 3 remains possible.
The fluctuations observed even for very high $p$ would be caused by
the logarithmic corrections present in model~II due to the
marginally irrelevant operator, as mentioned in
Section~\ref{sec:candidates}. This operator is expected to induce a
correction in $1/\log p$ (which is in principle computable exactly using
quantum field theory techniques, since it is universal;
progress on this will be reported elsewhere),
plus higher corrections, all of which remain significant in our range
of data.
This would also explain
why it is so hard to extract useful information from the first few
(exact) values of $a_p(n)$ given in \cite{jzja,jzjb}.

In conclusion, and in
view of the theoretical as well as numerical evidence, our belief is that
Conjecture 3 is indeed correct.

%Finally, we propose
%the following estimates for the 
%parameters of Eq.~(\ref{defk}): $\sigma' \doteq
%0.0480436$, $\gamma' \doteq 0.3$, $\kappa' \doteq -0.3$.

\section{Variants and corollaries}
First observe that planar maps have in general no symmetries. More
precisely the fraction of planar maps with $p$ edges that have a non
trivial automorphism group goes to zero exponentially fast under very
mild assumption on the family considered \cite{wormald}.  If this
(very plausible) property holds then a typical closed curve will be
obtained by closing $d$ different open curves, where $d$ is the degree
of the outer face. But the average degree of faces in any fixed
4-regular planar map is four. Thus the relation $a_p\sim4\alpha_p$.

Second let us give a property illustrating the importance of the
critical exponent $\gamma$ as opposed to the actual value of $\tau$.
A closed plane curve $C$ is said to be \emph{$\alpha$-separable}, for
$0<\alpha\leq1$ a constant, if there exist two simple points $x$ and
$y$ of $C$ such that $\Gamma\setminus\{x,y\}$ is not connected and
both connected components contain at least $p^\alpha$ crossings. The
pair $(x,y)$ is called a cut of $C$. In other terms, $C$ is
$\alpha$-separable if it is obtained by gluing the endpoints of two
big enough open plane curves (up to homeomorphisms of the sphere).
\begin{corollary}
  Assume Conjecture~\ref{main} is valid, and consider a uniform random
  closed plane curve $\Gamma_p$ with $p$ crossings. The probability
  that $\Gamma_p$ is 1-separable decays at least like $p^\gamma\doteq
  p^{-0.77}$. More generally, if
  $\alpha>1/(1-\gamma)=(7-\sqrt{13})/6\doteq 0.56$, the probability
  that $\Gamma_p$ is $\alpha$-separable goes to zero as $p$ goes to
  infinity.

  For comparison, $\gamma=-1/2$ and $1/(1-\gamma)=2/3$ for doodles,
  which are thus easier to separate.
\end{corollary}
Indeed let us compute the expected number of inequivalent cuts of a
closed plane curve with $p$ crossings. When considered up to
homeomorphisms of the sphere, close plane curves with a marked cut are
in one-to-one correspondence with pairs of open plane curves. Hence,
with a factor $p$ for the choice of infinite face,
\begin{equation}\label{proba}
p\cdot\sum_{p'=q}^{p-q}\frac{a_{p'}a_{p-p'}}{\alpha_p}
<cst\cdot p\cdot\sum_{p'=q}^{p-q}\frac{(p')^{\gamma-2}(p-p')^{\gamma-2}}{p^{\gamma-2}}
=O({p}{q^{\gamma-1}}).
\end{equation}
In particular if $q\gg p^{1/(1-\gamma)}$ this expectation goes to zero
as $p$ goes to infinity. 

It is typical that in the computation of probabilistic quantities,
like in Equation~(\ref{proba}), the exponential growth factors cancel,
leading to behaviors that are driven by polynomial exponents. This
explains the interest of in these \emph{critical exponents} and gives
probabilistic meaning to their apparent \emph{universality}. As a
final illustration of this point let us present two variants of
Conjecture~\ref{main}: (Definitions of prime self-intersecting curves
and alternating knots can be found in \cite{jzja,kjs}.)
\begin{conjecture}\label{last}
  The number $\alpha'_p$ of closed prime self-intersecting curves
  with $p$ crossings and the number $\alpha''_p$ of prime
  alternating knots with
  $p$ crossings lie in the same universality class as closed
  self-intersecting curves: there are constants $\tau'$, $\tau''$,
  $c'$, $c''$ such that
\[
\alpha'_p\sim c'\,\tau'{}^p\cdot p^{\gamma-2},%\cdot\log^\nu p,
\qquad
\alpha''_p\sim c''\,\tau''{}^p\cdot p^{\gamma-3},%\cdot\log^\nu p,
\]
where $\gamma$ is given in Conjecture~\ref{main}.
\end{conjecture}
Observe that knot diagrams are naturally considered up to
homeomorphisms of the sphere \cite{jzja,kjs}, while we have considered
plane curves up to homeomorphisms of the plane. This explains the
discrepancy of a factor $p$ in Conjecture~\ref{last} for $\alpha''_p$,
since one of the $p+2$ faces of a spherical diagram must be selected
to puncture the sphere and put the diagram in the plane.

\section{Conclusion}\label{sec:conclusion}
We have given arguments supporting Conjecture~\ref{main} for the
asymptotic number of plane curves with a large number of
self-intersections, as well as the more general
Conjecture~\ref{model2}. The numerical results provided in
Section~\ref{sec:simulations} support Conjecture~\ref{main} only
indirectly since they are related to another specialization of
Conjecture~\ref{model2} (derivative at $n=1$ versus $n=0$). However
the alternative proposal is not compatible with either of these new
numerical results (as is the case of Conjecture~\ref{model1}) or
earlier ones. 

Our method to test the conjecture could be applied to other models
like open curves with endpoints that are not constrained to stay in
the infinite face, or the meanders studied by Di Francesco et al.

\subsection*{Acknowledgements}
P.Z.-J. would like to thank J. Jacobsen for pointing out
Refs. \cite{jrs} and \cite{rs} to him.

\end{document}